\documentclass[fleqn,usenatbib]{mnras}

\usepackage{newtxtext,newtxmath}
\usepackage[T1]{fontenc}
\DeclareRobustCommand{\VAN}[3]{#2}
\let\VANthebibliography\thebibliography
\def\thebibliography{\DeclareRobustCommand{\VAN}[3]{##3}\VANthebibliography}
\usepackage{graphicx}	
\usepackage{amsmath}	

\title[]{Massive black hole binaries as sources of low-frequency gravitational waves and X-shape radio galaxies}

\author[Cury\l{}o \& Bulik]{
Ma\l{}gorzata Cury\l{}o$^{1}$\thanks{E-mail: mcurylo@astrouw.edu.pl},
Tomasz Bulik$^{1,2}$
\\
$^{1}$Astronomical Observatory, University of Warsaw, Al. Ujazdowskie 4, 00-478 Warsaw, Poland\\
$^{2}$Astrocent, Nicolaus Copernicus Astronomical Center, Rektorska 4, 00-614 Warsaw, Poland\\
}

\date{Accepted XXX. Received YYY; in original form ZZZ}

\pubyear{2023}

\begin{document}
\label{firstpage}
\pagerange{\pageref{firstpage}--\pageref{lastpage}}
\maketitle

\begin{abstract}
We present the study of multi-messenger signatures of massive black hole (MBH) binaries residing in the centres of galaxy merger remnants. In particular, we first focus on the gravitational wave background (GWB) produced by an ensemble of MBH binary inspirals in the frequency range probed by the Pulsar Timing Array (PTA) experiments. The improved estimates of the characteristic strain were obtained with the inclusion of environmental effects on the MBH binary orbital decay within the galaxy merger remnants, added in post-processing to the semi-analytic model of galaxy formation and evolution SHARK. Secondly, we explore two, intriguing in terms of the MBH binary evolution studies, hypotheses aiming to explain the origins of X-shape radio galaxies - a peculiar type of objects with double lobe structures, constituting approximately 6\,-\,10\% of known radio loud galaxies. The two considered scenarios involve either an abrupt change in the jet direction after a MBH merger (a spin-flip) or an unresolved close binary, where each of the two components produces a jet. We find that the estimated GWB amplitude at the reference frequency $f_0=1 \,{\rm yr}^{-1}$ is in the range of $A_{\rm{ yr^{-1}}} = 1.20\cdot10^{-15} - 1.46\cdot10^{-15}$, which is 50\% lower than the strain of the signal detected by the PTA experiments. We also show that the spin-flip scenario considered in gas-poor mergers reproduces the observed properties of X-shape radio galaxies well in terms of flip angle, redshift and luminosity distributions. 
\end{abstract}



\section{Introduction}
Massive black holes, with masses in the range of $\rm10^6~-~10^{10}\,M_{\odot}$, are one of the key building blocks in the hierarchical models of formation and evolution of galaxies since it was realised that their feedback is necessary to explain the bright end of galaxy luminosity functions \citep{Benson03, Croton06}. It was over a decade earlier when the first scaling relations linking them with their hosts were observed \citep{Dressler89} and they now constitute an extensive set of observational evidence for a mutual MBH and galaxy evolution. Within the most commonly brought up relations are those between the mass of the central MBH and galaxy stellar mass, luminosity and bulge velocity dispersion (\citealp{McConnell13, Kormendy13, Reines15, Gultekin19}, and references therein). 

In the hierarchical models, the build up of large scale structures occurs via consecutive mergers and matter accretion. It is therefore natural to expect the formation of MBH binaries in the centres of galaxy merger remnants \citep{Begelman80}, especially given the vast observational evidence for MBHs' presence in the bulges of both dwarf and large galaxies \citep{Ferrarese05, Vestergaard06, Reines13, Mezcua16}. Unfortunately, as opposed to a large collection of galaxy merger observations, close (less than a few pc apart) MBH binaries have not yet been detected, due to an insufficient resolving power of current telescopes. This raises the fundamental question whether such systems exist at all or, for instance, their orbital evolution is stalled at pc\,-\,kpc separations. 

In the pioneering work predicting the formation of MBH binaries, \cite{Begelman80} already delineated processes governing their orbital evolution. In the first phase soon after the hosts' merger, MBHs are still separated by $\sim$\,kpc and sink towards the centre of the remnants freshly settled core predominantly via dynamical friction. The next stage starts when the two MBHs form a gravitationally bound binary with a semi-major axis of the order of several pc. This phase is much more sensitive to the properties of the remnant galaxy and further binary evolution will depend on whether the environment is gas-rich or gas-poor. In the former case, large reservoirs of gas in the bulge result in the formation of a circumbinary disk around the MBHs which extracts their orbital energy and angular momentum by the viscous drag \citep{Cuadra09, Haiman09, Kocsis11, Roedig12, Roedig14_visc,Dotti15}, whereas binary hardening in gas-poor environments is dominated by scattering events with background stars \citep{Sesana06, Khan11, Khan13, Merritt13, Vasiliev15}. Finally, at sub-pc separations of the order of 0.001\,pc, the emission of gravitational waves takes over and quickly leads to the coalescence of the MBHs. Additionally, it has been also shown that multiple interactions with another MBH can efficiently bring the binary to mpc separations even if other processes fail to do so \citep{Hoffman07, Kulkarni12, Bonetti18}. Until today, the general idea of the binary evolution remains relatively unchanged (for a review see e.g. \citealp{Colpi14}), however the details of each process and the corresponding timescales are subject to extensive research. Clearly, one of the most crucial elements which are still missing is the direct observation of a statistically significant sample of binary systems at each evolutionary stage. 

Within the theoretical studies, there has been a lot of effort directed towards providing a variety of possible signatures of close MBH binaries, and they are being actively searched for in observational campaigns or archival data studies \citep{DeRosa19,Saade20,Bogdanovic22}. Some of the most prominent indications include: a) periodicity in emission due to jet precession, doppler boosts or self lensing \citep{DOrazio15, Haiman17, Dorazio18, Charisi18}, b) spectral features in light curves such as offset broad and double peaked narrow emission lines \citep{Bogdanovic08, Tsalmantza11}, and c) X-ray emission from circumbinary and mini-disks surrounding individual MBHs in the binary \citep{Komossa03, Sesana12, Roedig14_signa}. These endeavours resulted in the discovery of hundreds of candidate binary systems, however only several tens have been unambiguously confirmed with typical separations of the order of a few kpc (e.g. \citealp{Komossa03, Hudson06, Piconcelli10, Comerford15, Husemann18}, and others), and the tightest binary detected at separation of 7.3 pc \citep{Rodriguez06}. This is due to the fact that the proposed binary signatures can also be associated with other phenomena. For instance, even a direct radio image of an apparent dual core can in fact be related to a single active galactic nuclei (AGN) and a star forming region \citep{DeRosa19}. Confirmation of the MBH binary thus requires demanding multi-band observations and a clear evidence obtained with at least two separate methods. 

Lately, there has also been a growing interest in X-shape radio galaxies (XRGs) as potential hosts of MBH binaries or descendants of their mergers. These sources are characterised by an additional pair of lobes (the "wings") which are misaligned with respect to the primary ones and constitute approximately 6\,-\,10\% of know radio loud AGN (RLAGN; \citealp{Yang19}, however see also \citealp{Roberts15}). The number of known candidate XRGs has been relatively small, although recent new catalogue releases expanded them to the total number of over 600 \citep{Leahy92, Cheung07, Proctor11, Yang19, Bera20, Bera22, Bhukta22}.
There are several models attempting to explain the physical mechanism of XRG formation (see \citealp{Gopal-Krishna12, Lal19} for reviews) and the most promising include: 
\begin{itemize}
    \item backflow of plasma from the active to secondary lobes \citep{Leahy84, Capetti02, Hodges-Kluck11}
    \item rapid jet reorientation via Bardeen-Petterson effect \citep{Liu04}
    \item retrograde\,-\,prograde spin transition via Lense-Thirring effect \citep{Garofalo20}
    \item jet-shell interactions within the post-merger galaxy \citep{Gopal-Krishna10}
    \item spin-flip during a MBHB merger \citep{Merritt02, Dennett-Thorpe02, Gergely09}
    \item twin AGN with close unresolved MBH binary \citep{Lal05, Lal07}
\end{itemize}
Analyses of follow-up multi-band observations often
lead to contradictory conclusions and it has been suggested that there could be multiple avenues of XRG formation rather than one common process \citep{Lal19}. For instance, \citealp{Mezcua11} shows that XRGs have higher MBH masses and older starburst activities than AGN in a control sample, which supports the merger scenario. Similar conclusions were drawn from the observation of X-ray cavities in one of the XRG sources \citep{Hodges-Kluck10}. Moreover, several other studies  indicate a rapid spin reorientation on timescales of $\lesssim 10^5$\,yr \citep{Hernandez-Garcia17, Saripalli18, Bogdanovic22}. On the other hand, spectral analysis of several sources revealed that ages of primary and secondary lobes are similar so they cannot be associated with one jet which rapidly changed its direction and rather come from an unresolvable dual-jetted MBH binary system \citep{Lal05,Lal07}. Additionally, many XRGs do not show any signatures of past mergers and reside in relatively empty environments \citep{Landt10}, although note that minor mergers are usually hard to trace and, as \citealp{Rottmann02} pointed out, dense galaxy clusters enhance the fading time of the lobes which could explain higher probability of observing XRGs as rather isolated sources. 

Direct evidence that MBH do form close binaries and merge in less then the Hubble time would be provided by the detection of low-frequency gravitational waves (GWs). Assuming a circular binary motion, signal emission occurs at twice the orbital frequency, which for a MBH binary with a total mass of $10^6\,-\,10^{10}\,{\rm M_{\odot}}$ and $10^{-4}\,-\,1$ pc separation, would fall in the range of $10^{-10}\,-\,10^{-7}$~Hz. This is far below the frequencies probed by ground-based detectors LIGO/Virgo/Kagra, however it is exactly the target of the Pulsar Timing Array (PTA) experiments. The structure of PTA is based on four individual collaborations, namely the European PTA (EPTA; \citealp{Kramer13, epta23_data}), the North American Nanoherz Observatory for Gravitational Waves (NANOGrav; \citealp{Demorest13, Nano23_data}), the Indian PTA (InPTA; \citealp{Tarafdar22}) and the Parkes PTA (PPTA; \citealp{Manchester13,ppta23_data}), which jointly contribute to the International PTA (IPTA; \citealp{Hobbs10, Perera19}). 

Using decades of millisecond pulsar observations, PTAs search for 10s\,-\,100s nanosecond deviations induced by GWs passing between the Earth and the galactic distribution of pulsars in the meticulously measured pulse times of arrival. With current observational baselines of almost 20 years, PTAs are now beginning to be sensitive to GWB levels theoretically predicted for an ensemble of the cosmological population of inspiralling MBH binaries. Recently, all individual PTA collaborations (and later also the full IPTA) announced a detection of a common red process (CRP; \citealp{Arzoumanian20, Chen21, Goncharov21, Chen22}), however due to lack of the essential Hellings\&Downs correlations expected for a GWB \citep{Hellings83}, detection could not have been claimed as the signal may as well have a source in, e.g. an unknown noise process or a flaw in analysis methods \citep{Zic22}. The amplitude of the CRP signal, in the range of $A_{{\rm yr^{-1}}} = 1.95\,-\,2.95\cdot10^{-15}$ (as measured by individual PTAs and the IPTA at a reference frequency of $f_0 = {\rm yr^{-1}}$) was broadly compared with predictions based on theoretical models and EM observations to asses whether the detected signal could indeed be of astrophysical origin. Unsurprisingly, estimates cover a wide range of amplitudes from $10^{-16}$ up to $10^{-14}$ for the most optimistic models \citep{Rosado15, Middleton21}, because of the large number of processes and unknowns involved in the cosmological evolution of MBHs and their host galaxies. On the other hand, however, \cite{Izquierdo-Villalba22} points out that higher GWB amplitudes consistent with the CRP can be reproduced by enhancing MBH growth which in turn introduces significant discrepancy with electromagnetic (EM) constraints, such as quasar bolometric luminosity and the local BH mass functions. 

Following the detection of the CRP, new extended data sets from all major PTAs along with the Chinese PTA (CPTA) provide the first in history evidence that the signal is consistent with the GWB \citep{EPTA23, Nano23,PPTA23, CPTA23}. 
However, due to still low significance (between 2$\mathrm{\sigma}$ and 4$\mathrm{\sigma}$) definite confirmation of the source and characterisation of the signal will require further years of observations and study. In particular, a new combined data set of the whole IPTA is expected to reveal more details because of its higher sensitivity with respect to individual PTAs \citep{ipta23_comp}.

In the following paper we present the study of binary MBHs and their mergers in terms of their possible GW and EM emission and associate them with their host galaxy properties. Our work is based on the results of the semi-analytic model of galaxy formation SHARK \citep{Lagos18} which provided us with an initial sample of galaxy mergers hosting MBHs at their centres (Sec.\,\ref{sec:population}). We first started with modelling MBH binary evolution within galaxy merger remnants in order to estimate how many will form bound systems and coalesce within the Hubble time (Sec.\,\ref{sec:binary_evolution}). Further, we used these evolved binaries to calculate the amplitude of the GWB and compare it with most recent PTA results (Sec.\,\ref{sec:lifetimes}). In the second part of the paper we focus on the EM emission of our evolved binaries, specifically related to their jet morphologies. For the first time, we used large-scale cosmological simulations to study whether MBH binaries and their mergers can be progenitors of double-lobe structured X-shape radio galaxies (Sec.\,\ref{sec:xrgs}). Throughout the paper, we use cosmological constants from \cite{Planck16}: $h\,=\,0.6751$, $\Omega_{\rm m}=0.3121$, $\Omega_{\Lambda}=0.6879$.

\section{MBH binary population}\label{sec:population}
In order to get our initial MBH binary population, we used the semi-analytic model of galaxy formation and evolution SHARK \citep{Lagos18}. The model is built upon DM halo catalogues and merger trees from the SURFs N-body simulations \citep{Elahi18}, which are available in four volume/resolution variants. In this work, we used simulations with a box size of $\rm(210\,cMpc/h)^3$, $1536^3$ particles and softening length of 4.5 ckpc/h. 

Galaxy evolution is based on a set of analytical prescriptions of all major processes, such as gas cooling/accretion, star formation, stellar feedback and reionization, which can be chosen from a relatively rich build-in library (for details, see \citealp{Lagos18}). In the default model, which we employed for our analysis, each DM halo with $\rm m_{halo}>10^{10}\,M_{\odot}$ is seeded with a central MBH with $\rm m_{seed} = 10^4\,M_{\odot}$. Subsequent growth of MBHs can occur via three channels, 1) mergers and accretion in 2) radio or 3) quasar modes. The latter corresponds to starburts, which are triggered by both mergers and disk instabilities and at the same time plays the dominant role in the overall growth of the MBHs. On the other hand, during radio mode MBHs grow via Bondi-Hoyle like accretion and are a source of AGN feedback through relativistic jets. Spin evolution of the MBHs is not traced and instead, a constant radiative efficiency $\epsilon = 0.1$ is applied (corresponding to a spin parameter $a \sim 0.67$). Finally, the output of the simulations is written out in 200 snapshots covering the evolution in the redshift range of $z = 24 - 0$. 

For the purposes of our analysis, we extracted information of all merging galaxies that contain MBHs at their centres and in order to avoid populating our binary sample with sources that may likely be a result of numerical errors or modelling imperfections, we further utilised the following criteria:
\begin{itemize}
    \item All MBHs must be present in the snapshot prior to the merger.
    This allows us to exclude low-mass seeds merging right after they are created. A similar cut-off was employed in \cite{Kelley17}. 
    \item MBHs have a non-zero total accretion rate.
    \item Galaxy total stellar mass is not lower than the resolution of the simulations $M_* > 10^7\,{\rm M_{\odot}}$. 
\end{itemize}
After this filtering procedure, we were left with $\sim$89\% of the initial data set and note that most of the sources were excluded based on the first criterion.


\section{Binary evolution}\label{sec:binary_evolution}
Mergers in SHARK occur after DM halo collisions. Galaxy which resided in the main progenitor halo is redefined as the central one and the remaining galaxies become satellites that sink towards the centre of the halo via dynamical friction in time given by \cite{Lacey93}:
\begin{equation}
    \tau_{\rm galaxy} = f_{\rm df}\,\Theta_{\rm orb}\,\tau_{\rm d}\,\left(\frac{0.3722}{\ln(\Lambda_{\rm Coulomb})}\right)\,\frac{M}{M_{\rm sat}}
\end{equation}
where $f_{\rm df} \le 1 $ is a free parameter, $\tau_{\rm d} = R_{\rm V}/V_{\rm V}$ is the dynamical timescale of the halo, $\ln(\Lambda_{\rm Coulomb})= \ln(M/M_{\rm sat})$ is the Coulomb logarithm, $M$ is the halo mass of the central galaxy, $M_{\rm sat}$ is the mass of the satellite (including its DM halo) and $\Theta_{\rm orb}$ is an orbital function whose value is drawn from a log normal distribution with a median $\log_{10}(\Theta_{\rm orb}) = -0.14$ and dispersion $\sigma = 0.26$. 

After $\tau_{\rm galaxy}$ SHARK galaxies merge and so do their central MBHs. However, this simple model does not include any estimation of the time that is needed for the MBHs to actually reach the centre of the remnant galaxy, form a gravitationally bound binary and coalesce. 

Below, we present modelling of the binary evolution added in post-processing the MBH binary population retrieved from the simulation at the time of their host galaxy merger.

In summary, we model the binary orbital decay as dependent on the parameters of the remnant galaxy and driven by four main processes:
\begin{itemize}
    \item dynamical friction within the remnant
    \item stellar scattering
    \item viscous drag from circumbinary disk
    \item GWs.
\end{itemize}

\subsection{Dynamical friction}

After the merger of galaxies, central MBH sinks towards the centre of the remnant galaxy via dynamical friction from interactions with stars. The process reduces the separation of the MBHs from $\sim$kpc to $\sim$pc scales, where they form a gravitationally bound binary system. 

We calculate the dynamical friction timescale for the secondary MBH (assuming the primary to be located at the centre) following \cite{Binney08}:
\begin{equation}\label{eq:df}
    t_{\rm dyn} = 19\,\left(\frac{r_0}{5\,kpc}\right)^2\,\left(\frac{\sigma}{200\,{\rm km\,s^{-1}}}\right)\,\left(\frac{10^8\,{\rm M_{\odot}}}{M_{2}}\right)\,\frac{1}{\Lambda}\,[{\rm Gyr}]
\end{equation}
where $r_0$ is the initial MBHs separation (defined as the half-mass radius of the remnant galaxy), $\sigma$ is the velocity dispersion of the remnant galaxy, $\Lambda = \ln(1+M_*/M_{2})$ is the Coulomb logarithm and $M_*$ and $M_{2}$ are the total stellar mass of the remnant galaxy and the mass of the secondary MBH, respectively. 

 We assume that MBHs form a bound binary at a separation which we set to $a_0\,=\,GM/2\sigma^2$, where $M$ is the total mass of the binary and $G$ is the gravitational constant. 

\subsection{Stellar scattering and viscous drag from circumbinary disk}

Further evolution of the binary will depend on the properties of merging galaxies. We can divide mergers into two groups, 1) gas-rich (wet) mergers, when gas reservoir in the bulge is larger than the mass of the binary (these constitute $\sim$69\% of the our data set) and 2) gas-poor (dry) mergers, when gas supplies are small.

In the case of gas-rich mergers the binary semi-major axis is shrinking due to interaction with the circumbinary disk. We use a simple model of the viscous drag exerted by the disk from \cite{Dotti15} (see also \citealp{Bonetti19, Volonteri20}):
\begin{equation}
    t_{\rm gas} = 1.5 \cdot 10^{-2}\,f_{\rm edd}^{-1}\,\frac{q}{(1+q)^2}\,\ln\left(\frac{a_0}{a_{\rm c}}\right)\,[{\rm Gyr}]
\end{equation}
where $f_{\rm edd}$ is the Eddington ratio (luminosity in units of Eddington luminosity), $q = M_2/M_1 \leq 1$ is the MBHs mass ratio and 
\begin{equation}
    a_{\rm c}\,=\,1.9\cdot10^{-3}\,\left(\frac{M}{10^8\,{\rm M_{\odot}}}\right)^{3/4}\,[{\rm pc}]
\end{equation}
is a conservative estimate of the separation at which GWs alone can bring the MBH binary to coalescence in less than the Hubble time ($\sim10^5$\,yr in the case of an equal mass binary). The $f_{\rm edd}$ should correspond to the total accretion rate for both MBHs in the binary, hence we use rate of the remnant MBH saved at the snapshot directly following the merger.  

In the case of gas-poor mergers, the orbital energy and angular momentum are carried away by numerous scattering events with background stars and the efficiency of the process depends on the density of stars in the so called "loss-cone" (i.e. objects with angular momentum small enough in order to come into interaction with the MBH binary). However, in the course of the binary hardening, the number of stars may significantly drop, leading to a stagnation of the orbital decay. Several processes of the cone replenishment have been proposed, although the details and actual potency in reducing the binary separation with interactions with stars remain as open questions \citep{Merritt13}. In this work, we assume that the loss-cone is refilled efficiently (the "full loss-cone") and calculate the hardening timescale following \cite{Sesana15} and \cite{Volonteri20}:

\begin{equation}
    t_{\rm star} = 15.18\,\left(\frac{\sigma_{\rm inf}}{\rm km\,s^{-1}}\right)\,\left(\frac{\rho_{\rm inf}}{\rm M_{\odot}pc^{-3}}\right)\,\left(\frac{a_{\rm GW}}{10^{-3} {\rm pc}}\right)\,[{\rm Gyr}]
\end{equation}
where $\sigma_{\rm inf}$ and $\rho_{\rm inf}$ are velocity dispersion and the stellar density at the radius of influence given by \citep{Volonteri20}:

\begin{equation}
    \begin{split}
        r_{\rm inf}    &= R_{\rm eff} \left(\frac{4M}{M_*}\right)^{1/(3-\gamma)} \\
        \rho_{\rm inf} &= \frac{(3-\gamma)M_*}{8\pi} \frac{r_{\rm inf}^{-\gamma}}{R_{\rm eff}^{3-\gamma}}
    \end{split}
\end{equation}
where $R_{\rm eff}$ is the effective radius of the remnant galaxy, $\gamma\,=\,1.75$ is the index of the stellar density profile and $a_{\rm GW}$ corresponds to the separation at which binary stays for the most of the time (hardening processes are least efficient):

\begin{align}
    a_{\rm GW} = & \,2.64 \times 10^{-2} \nonumber \\
    & \times \left[ \left(\frac{\sigma_{\rm inf}}{\rm km\,s^{-1}}\right)\,\left(\frac{\rm M_{\odot}pc^{-3}}{\rho_{\rm inf}}\right)\,\left(\frac{15}{H}\right)\,\left(\frac{M_1M_2M}{2\cdot10^{24} \rm M_{\odot}^3}\right) \right]^{1/5}\,[{\rm pc}]
\end{align}

We follow \cite{Volonteri20} and adopt H=15, which is a dimensionless hardening rate estimated from scattering experiments \citep{Quinlan96, Sesana15}. 

\begin{figure}
    \centering
    \includegraphics[width=\columnwidth]{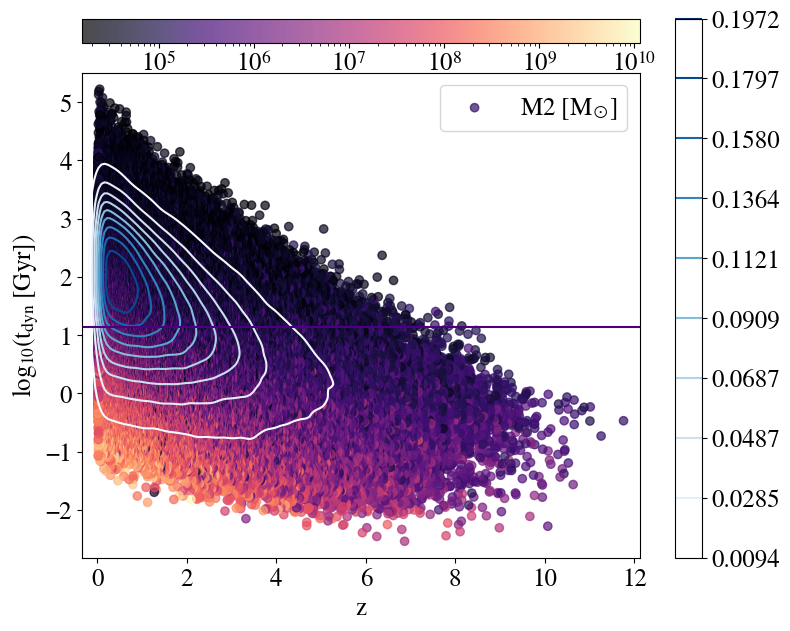}
    \includegraphics[width=\columnwidth]{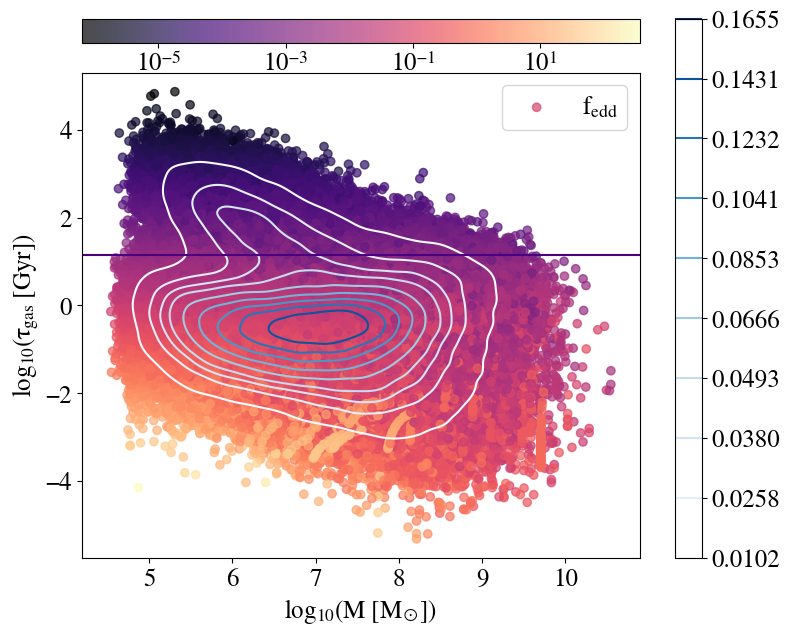}
    \includegraphics[width=\columnwidth]{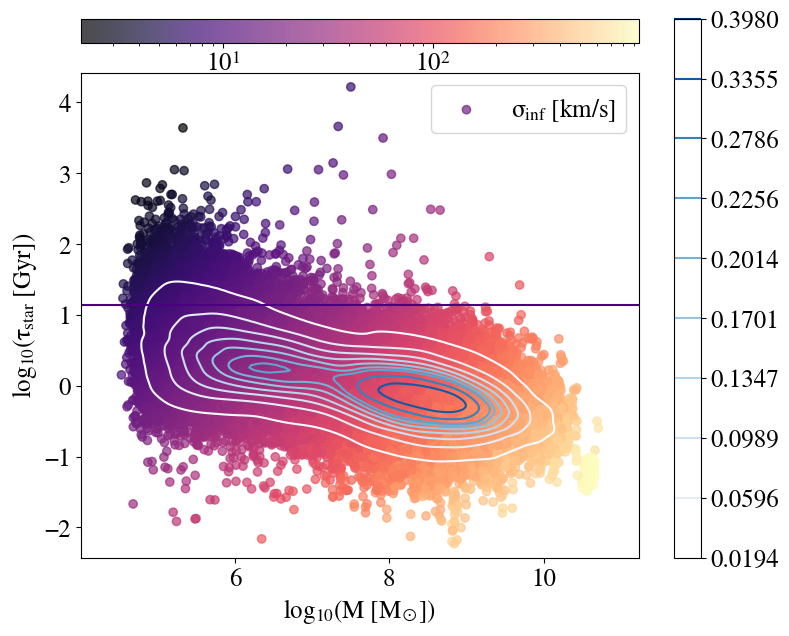}
    \caption{Delay timescales due to dynamical friction as a function of redshift, and viscous drag and stellar scattering as a function of total binary mass $M$, with colour bars representing the secondary MBH mass $M_2$, Eddington ratio $f_{\rm edd}$ and velocity dispersion $\sigma_{\rm inf}$. The blue horizontal line in each panel references the Hubble time. Additionally, on each panel we also include normalised density contours to better characterise the distributions.} 
    \label{fig:df}
\end{figure}


Fig.\,\ref{fig:df} shows the delay timescales due to dynamical friction, viscous drag and stellar scattering. If we recall Eq.\, \ref{eq:df}, we can see that $\tau_{\rm dyn}$ has a strong dependence on initial MBHs separations which are naturally much smaller at high redshifts. As a result, practically all MBHs at $z \ge 8$ and in total $\sim$40\% of the whole data set will form a bound binary in less than the Hubble time. As for the further evolution of gravitationally bound binaries, the assumed full loss-cone model for stellar scattering results in 97\% of gas-poor mergers having hardening timescales smaller than the Hubble time, while for gas-rich mergers and orbital decay due to viscous drag it is 88\%. The actual number of coalescences in each case is presented in the Section \ref{sec:lifetimes}, where the new MBH merger redshifts were recalculated including the delay timescales.

\subsection{Gravitational waves}
The last stage of orbital decay, whether binaries evolve in gas-rich or gas-poor environment, is dominated by the emission of GWs. Assuming circular orbits, we can calculate the time to coalescence as:
\begin{equation}
    t_{\rm GW} = \frac{5c^5}{64G^3}\,\frac{a^4 - a_{\rm f}^4}{M_1M_2M}
\end{equation}
where $a_{\rm f} = 6\,GM/c^2$ is the innermost stable circular orbit (ISCO) of a Schwarzschild BH and $a$ is the separation at which GWs start to dominate the orbital decay set to $a = a_{\rm c}$.

In the calculations of the GW emission we did not take into account two factors. Firstly, we ignored non-zero eccentricities and their evolution, and secondly, we "turned off" all other hardening mechanisms when GW started to dominate. In the latter case, the GWB is attenuated at the lower frequency end (below $10^{-8}$\,Hz), while increasing eccentricity rises the amplitude of the characteristic strain above $10^{-8}$\,Hz as GW energy is distributed to higher harmonics (see \citealp{Kelley17}).

\section{Binary lifetimes and GWB estimates}\label{sec:lifetimes}

Accounting just for the dynamical friction within the remnant galaxy drastically reduces the number of MBHs that can effectively form a gravitationally bound system in less than the Hubble time. In Tab.\,\ref{tab:delay_numbers} we compare the number of binaries at each evolutionary stage. The last row shows the percentage of binaries that will merge at $z_{\rm merge}~\ge~0$ calculated after accounting for all delay mechanisms described in Sec. \ref{sec:binary_evolution} as:

\begin{equation}
    z_{\rm merge} = z(t_{\rm galaxy} + t_{\rm dyn} + t_{\rm stars/gas} + t_{\rm gw})
\end{equation}\\
where $z(t_{\rm galaxy})$ is the redshift of the original galaxy merger returned by SHARK.

\begin{table}
	\centering
	\caption{Percentage of binaries at each evolutionary stage relative to all galaxy mergers ("gal", second column) or binary systems ("bin", third column). Configurations: initial refers to initial filtering procedure, binary shows the number of bound binaries, and mergers correspond to binaries that merge before $z=0$. In summary, 40\% of galaxy mergers result in formation of a binary, 66\% of binaries will merge before z=0, which corresponds to 26\% of MBH coalescences out of all galaxy mergers.}
	\label{tab:delay_numbers}
	\begin{tabular}{l c c c} 
		\hline
		  Configuration      &  \% gal & \% bin  & Criteria    \\
		\hline
		All galaxy mergers & 100    & -       & Merging galaxies  \\ \\
		Initial            & 89     & -       & $M_{\rm p}>10^4\,{\rm M_{\odot}}$    \\
		Binary             & 40     & 100     & $\tau_{\rm dyn} < t_{\rm H}$    \\
        Mergers            & 26     & 66      & $z_{\rm merge} \ge  0$   \\
		\hline
	\end{tabular}
\end{table}

\begin{figure}
	\includegraphics[width=\columnwidth]{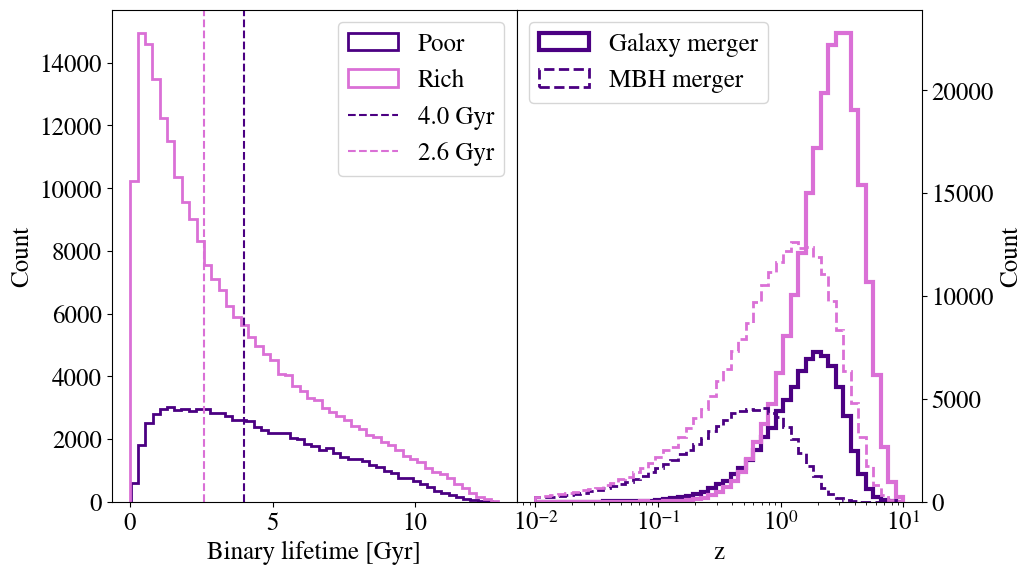}
    \caption{Binary lifetimes in gas-poor/gas-rich mergers and a comparison between galaxy merger redshifts (thick lines) with the delayed MBH merger redshifts (dashed lines). The two vertical lines on the left panel show the median values for both distributions.}  
    \label{fig:lifetimes}
\end{figure}

In Fig.\,\ref{fig:lifetimes} we show the distribution of binary lifetimes (defined as a sum of all delay times) for merging systems and a comparison of galaxy and MBH merger redshifts. Median values of the gas-rich (gas-poor) distributions are located at 2.6 Gyr (4.0 Gyr), and in general 22\% (9\%) of binaries merge within 1 Gyr. Median lifetimes of all binaries including stalled ones reach 4.5 Gyr and 6.6 Gyr, respectively. We can thus see that in our simulations, interaction with gas leads to quicker binary evolution than stellar scattering, which is a result of high accretion rates that can exceed Eddington limit and many low-mass ratio binaries. In the case of stellar scattering, the timescale strongly depends on the galaxy density profile characterised by the spectral index in the range of $0.5 < \gamma < 2$. Larger $\gamma$ corresponds to higher densities and thus shorter hardening timescales.
We chose $\gamma=1.75$ so that our model reproduces densities observed in local galaxies \citep{Pfister20}. For comparison, setting $\gamma=2$ would result in 99\% of stellar hardening timescales below the Hubble time, whereas for $\gamma=1$ it would drop to only 61\%.  

Finally, in order to calculate the characteristic strain of the GWB $h_{c}$ we consider only the binaries which merge before $z=0$ and assume that they are circular and GW-driven. Following \cite{Sesana08} we start with:

\begin{equation}\label{eq:gwb}
    h_{\rm c}^2(f) = \int_{0}^{\infty} {\rm d}z {\rm d}\mathcal{M} \frac{{\rm d}^3N}{{\rm d}z {\rm d}\mathcal{M} {\rm dln}f_{\rm r}} h^2(f_{\rm r})
\end{equation}
where $d^3N/dzd\mathcal{M}dlnf_{\rm r}$ is the comoving number of sources emitting in a logarithmic frequency interval per unit redshift and chirp mass $\mathcal{M}$, f and $f_{\rm r}$ are the observed and rest-frame GW frequencies and:
\begin{equation}
    h = \frac{8\pi^{2/3}}{10^{1/2}}\,\frac{(G\mathcal{M})^{5/3}}{d_{\rm c}(z)c^4}\,f_{\rm r}^{2/3}
\end{equation}
is the spectral strain of a single circular binary with $d_{\rm c}$ being the comoving distance. 
Further, we can expand the term for the comoving number of sources as \citep{Sesana08}: 
\begin{equation}
    \frac{{\rm d}^3 N}{{\rm d}z {\rm d}\mathcal{M} {\rm dln} f_{r}} = \frac{{\rm d}^2 n}{{\rm d}z {\rm d}\mathcal{M}} \frac{{\rm d}t_{\rm r}}{{\rm dln} f_{\rm r}} \frac{{\rm d}z}{{\rm d}t_{\rm r}} \frac{{\rm d} V_{\rm c}}{{\rm d}z}
\end{equation}
where $V_{\rm c}$ is the comoving volume and thus \citep{Hogg99}:
\begin{equation}
    \frac{{\rm d} V_{\rm c}}{{\rm d}z} \frac{{\rm d}z}{{\rm d}t_{\rm r}} = 4\pi c d_{\rm com}^2 (1+z)
\end{equation}
and: 
\begin{equation}
     \frac{{\rm d}t_{\rm r}}{{\rm dln} f_{r}} = \frac{f_{\rm r}} {{\rm d}f_{\rm r}/{\rm d}t_{\rm r}} = \frac{5}{96} \frac{c^5}{ (G\mathcal{M})^{5/3} } (\pi f_{\rm r} )^{-8/3}
\end{equation}
is the hardening timescale assuming a circular, GW-driven binary \citep{Kelley17}. 
Using the above relations we can rewrite Eq.\,\ref{eq:gwb} as:
\begin{equation}
     h_{\rm c}^2(f) = \frac{20\pi c^6}{96} \int_0^{\infty} {\rm d}\mathcal{M} {\rm d}z \frac{{\rm d}^2 n}{{\rm d}z {\rm d}\mathcal{M}} h^2(f_{\rm r}) \frac{d^2_{\rm c}(1+z)}{(G\mathcal{M})^{5/3}} (\pi f_{\rm r})^{-8/3}
\end{equation}
As we are using a finite number of binaries returned by the simulation, we can replace the integral by a sum over all sources in the simulated comoving volume $V_{\rm sim}$ emitting at each PTA frequency:
\begin{equation}
    \int {\rm d}\mathcal{M} {\rm d}z \frac{{\rm d}^2 n}{{\rm d}z {\rm d}\mathcal{M}} ... \rightarrow \frac{1}{V_{\rm sim}} \sum_i ...
\end{equation}
Finally, it is convenient to express the characteristic strain as:
\begin{equation}
    h_{\rm c}(f) = A_{\rm yr^{-1}}\left(\frac{f}{f_0}\right)^{-2/3} 
\end{equation}
where $A_{\rm yr^{-1}}$ is the GWB amplitude at a reference frequency $f_0 = yr^{-1}$.

\begin{figure}
	\includegraphics[width=\columnwidth]{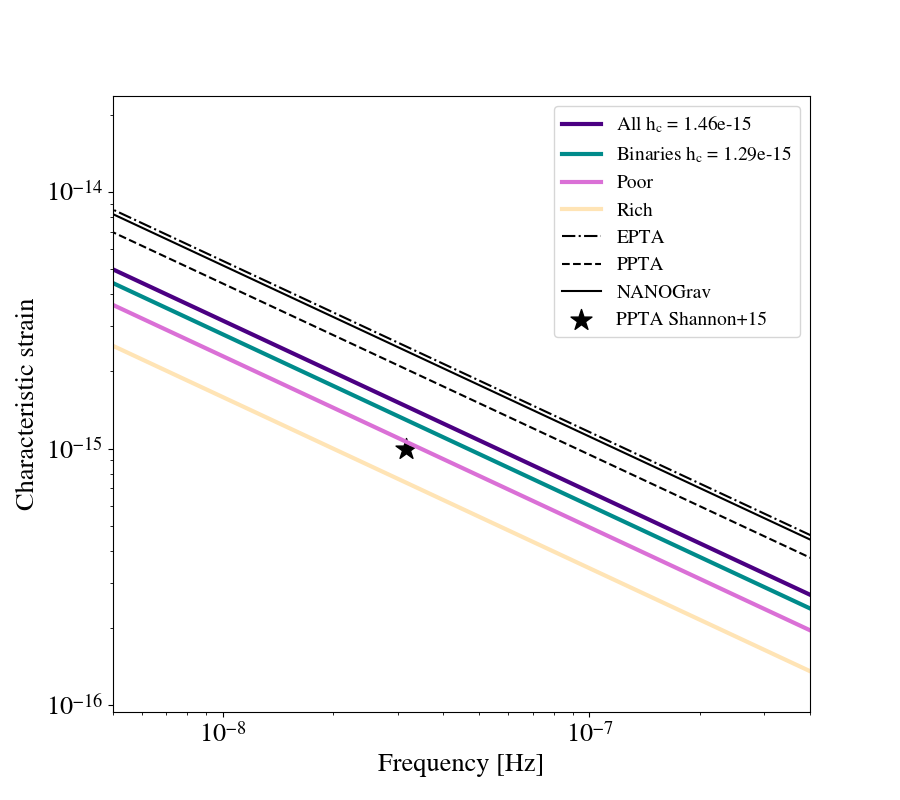}
    \caption{Characteristic strain of the GWB estimated in our calculations (thick colour lines) compared with observational constrains (black lines). The dark purple ("All") corresponds to all MBHB before including binary evolution models (first row in Tab.\,\ref{tab:delay_numbers}), while dark cyan shows only those binaries which merge before $z=0$. The black star shows the most stringent constraint from before CRP detection \citep{Shannon15}.}
    \label{fig:GWB}
\end{figure}

Fig.\,\ref{fig:GWB} shows the GWB we calculated for all, merging and poor/rich binaries and compares them with observational constraints provided by all PTA groups. The predicted strain for all binaries (no-delays, all galaxy mergers result in MBH coalescence) and those which will merge after including all delay processes has a very similar amplitude ($A_{\rm yr^{-1}} = 1.29\cdot10^{-15}\,-\,1.46\cdot10^{-15}$). This comes from the fact that most of the signal in the PTA range comes from the most massive ($10^7\,{\rm M_{\odot}}$) and relatively close by ($z < 2$) sources, which also have short binary lifetimes. Additionally, out of all binaries produced in SHARK, only 11\% significantly contribute to the GWB in the PTA range. 
These results were obtained for our fiducial stellar density profiles with $\gamma=1.75$. However, the main contributors to the GWB at PTA frequencies are very massive MBHs which usually reside in galaxies characterised by shallower density profiles. We therefore also examined the case for $\gamma=1$ and the resulting GWB amplitude is only slightly smaller $A_{\rm yr^{-1}} = 1.20\cdot10^{-15}$. This is again due to the fact that most massive MBHs have short delay times in both models of stellar scattering.

\begin{table}
	\centering
	\caption{Amplitudes of the GWB calculated in this work in comparison with observational constraints from EPTA, NANOGrav and PPTA. We also show example predictions obtained with two different methods: semi-analytic model L-Galaxies and hydrodynamic simulations Illustris with binary treatment added in post-processing. Whenever a range is given, it corresponds to the lower and upper limits when analysing delayed or non-delayed (all MBHs merge) models, respectively.}
	\label{tab:gwb_amplitudes}
	\begin{tabular}{l c c } 
    \hline
    PTA        & GWB $A_{\rm yr^{-1}} / 10^{-15} $                    & Reference \\
    \hline\hline
    EPTA       & $2.5\pm0.7 $           & \cite{EPTA23}    \\ \\
    NANOGrav   & $2.4^{+0.7}_{-0.6}$    & \cite{Nano23} \\ \\
    PPTA       & $2.0^{+0.3}_{-0.2}$    & \cite{PPTA23} \\ \\  
    L-Galaxies & 1.2                    & \cite{Izquierdo-Villalba22} \\ \\
    Illustris  & $0.4 - 0.7$            & \cite{Kelley17}  \\ \\
    This work  & $1.2 - 1.5$            &              \\
 
    \hline
	\end{tabular}
\end{table}

Most importantly, even assuming that all MBHs coalesce immediately after the merger of their host galaxies, the calculated GBW is 50\% lower than the signal detected by the PTAs. The simplest explanation in favour of the SHARK galaxy model would be that the PTA signal is not in fact produced by MBH mergers. First of all, it has been shown to be consistent also with other sources of GWs \citep{Afzal23, EPTA_astro} and secondly, the detection evidence itself still has a relatively low significance. Nevertheless, MBHs are still prime sources of the GWB and it is not unlikely that in the next few years the IPTA combined analyses will be able to confirm it, especially that some reasonably parametrised models can reproduce the signals amplitude \citep{Middleton21, Agazie23, EPTA_astro}. It is therefore most sensible to reevaluate our simulations and answer the question if it is possible to increase the calculated GWB strain providing that the model is reproducing other observational constraints well or, in other words, to treat the PTA signal as a new constraint to be included in the model. Practical investigation is out of the scope of this paper, however there are a few ideas that could be explored in future work. Firstly, MBH binary evolution could be solved within the simulations instead of being done in post processing. As we have shown, only 26\% of all galaxy mergers result in subsequent coalescence of the MBHs and 40\% create a bound binary (note however that we calculated the DF timescale for the less massive MBH, while a primary may reach the galactic centre faster). This can have substantial impact on the MBH mass and luminosity functions, AGN feedback and the overall galaxy population. Noteworthy, the coalescence fractions can, in fact, be higher after including multiple MBH interactions and eccentricity to the evolution models, which were omitted in our analysis. Approximately 8\% of all galaxy mergers in our model are followed by at least one more merger within the same snapshot, and additionally, accounting for long binary lifetimes we can expect the percentage of multiple interactions to increase. However, as \cite{Bonetti18} showed, stellar and gas interactions are still dominant processes of binary evolution. Finally, \cite{Siwek20} showed that allowing for accretion during binary migration from kpc separations down to the coalescence can increase the GWB amplitude even by a factor of 4. 

Nevertheless, what is worth noticing is that most of the self-consistent models of mutual galaxy and MBH evolution which are in agreement with EM observational constraints predict GWB amplitude at the level of $A_{\rm yr^{-1}} \sim 1\cdot10^{-15}$ (e.g. \citealp{Barausse20, Siwek20, Izquierdo-Villalba22}). There is therefore some tension between the most recent PTA results and our current understanding of galaxy evolution which will drive exciting research probably for the next decades.

\section{X-shaped galaxies as signatures of binary MBHs}\label{sec:xrgs}
In the following sections we will investigate whether binary MBHs and their mergers can be progenitors of an interesting class of objects: the X-shaped radio galaxies. From now on, we will only use keplerian binaries, i.e. MBHs with $\tau_{\rm df} < t_{\rm H}$ either stalled at few pc separations or those which will merge before $z=0$ (denoted as "binary" in Tab.\,\ref{tab:delay_numbers}). 

We start with selecting MBHs that are capable of producing jets based on observational constraints available in the literature. Next, we explore two configurations that can result in formation of an XRG, searching in our sample for: 1) MBH binaries with one jet and calculate its change of direction after a merger (the spin-flip scenario) and 2) dual-jetted AGN, i.e. MBH binary with two active MBHs with misaligned jets (the twin AGN scenario). In order to compare our estimates with XRG observations, we calculate the final flip angles and angles between the jets of the twin-AGNs including projection effects (each plane with intrinsic angle is randomly rotated; for details of the procedure see \citealp{Rottmann02}).

\subsection{Radio loud galaxies and jets: initial sample}
The processes of formation and propagation of astrophysical jets are still debated, despite the fact that they have been first discovered over a century ago \citep{Blandford19}. In order to estimate how many MBH in our data set can produce jets we use the correlations between radio loudness parameter (defined as ratio of radio and optical luminosities), Eddington ratio and MBH mass \citep{Ho02, Sikora07, Thomas21}. Additionally, we also take into account studies based on optical spectra of RLAGN, which relate the dominant mode of accretion onto the MBH with presence of high and low excitation lines. The former case occurs during the quasar mode, i.e. accretion of cold gas via geometrically thin disk and such objects are referred to as high excitation radio galaxies (HERGs). They are characterised by a broad EM emission and only 10\% of them produce large scale jets. Low excitation radio galaxies (LERGs) are powered by radiatively inefficient accretion during the radio mode and release most of their energy through relativistic jets. Noteworthy, it has been shown that LERGs are also associated with more massive MBHs and red ellipticals, while HERGs tend to show higher star formation rates (for a thorough review on AGN see \citealp{Tadhunter16}).
In each set of galaxies from gas-rich and gas-poor mergers we therefore search for radio loud sources with jets initially defined as LERGs, i.e. those where radio mode is the dominant mode of accretion and with $M>10^8~{\rm M_{\odot}}$ and $f_{\rm edd} < 0.02$. We will ignore possible jets in the small fraction of HERGs in the course of this work. We argument that choice by their small number and also the fact that over 90\% of XRGs are associated with Fanaroff-Riley class II galaxies (FRII; \citealp{Fanaroff74}) which in most cases are also found to be LERGs \citep{Capetti17}.

\begin{figure}
	\includegraphics[width=\columnwidth]{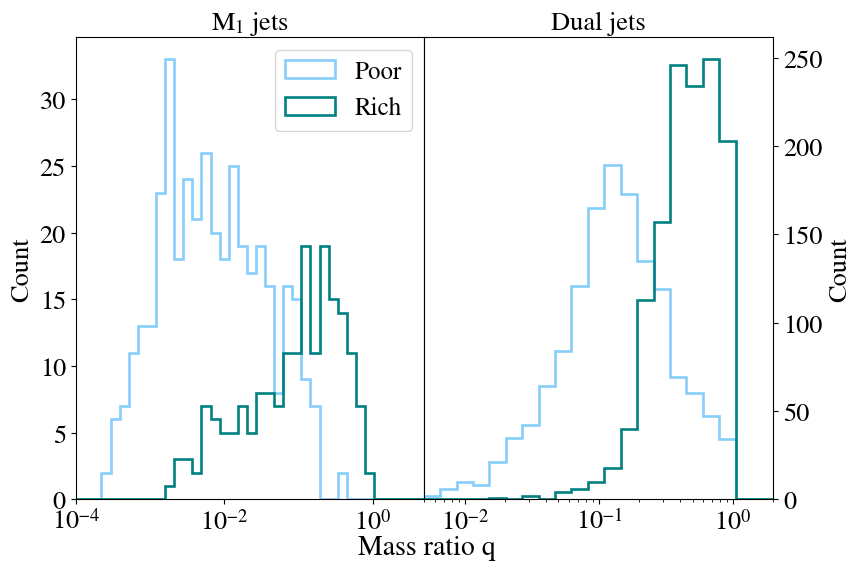}
    \caption{Distribution of mass ratios for single dual dual-jetted AGN in both gas-rich and gas-poor mergers.}
    \label{fig:qs}
\end{figure}

In Fig.\,\ref{fig:qs} we show the distributions of mass ratios. 
Gas-poor mergers rarely have MBH binaries with near equal masses, whereas gas-rich mergers peak at $q>0.1$. The initial sample of dual-jetted AGN is characterised by notably higher q and this will have a major impact on our final sample of such sources (discussed in more detail in Sec.\,\ref{subsec:dual} and \ref{subsec:xrg_results}).

\subsection{Spin-flip}
In the spin-flip scenario, direction of a jet produced by one of the MBHs in a binary (presumably the primary) will change its direction as a result of a merger. To make our calculations, we follow formulas presented in \cite{Hofmann16}. In particular, we assume that the final spin is parallel to the total angular momentum of the binary given by:
\begin{equation}
        J = S_1 + S_2 + L
\end{equation}
where $S_{1,2} = a_{1,2}\,GM_{1,2}/c^2$ are the individual MBH spins and $L = L_{\rm ISCO}(a)$ is the specific angular momentum (divided by $GM_1M_2/c$) at the ISCO of a Kerr BH:

\begin{equation}
   \begin{split}
    L_{\rm ISCO}(a) &= \frac{2}{3\sqrt{3}} \left[ 1+2\,\sqrt{3\,r_{\rm ISCO}(a) - 2}  \right] \\
    r_{\rm ISCO}(a) &= 3 + Z_2 \pm \sqrt{ (3 - Z_1)\,(3 + Z_1 + 2\,Z_2) } \\
    Z_1         &= 1 + (1+a^2)^{1/3} \left[ (1+a)^{1/3}+ (1-a)^{1/3} \right] \\
    Z_2         &= \sqrt{ 3\,a^2 + Z_1^2 } 
   \end{split}
\end{equation}
The minus and plus signs correspond to prograde and retrograde orbits, respectively and $a = a_{\rm tot}$ is the total spin parameter defined as:
\begin{equation}
     a_{\rm tot} (\beta, \gamma, q) = \frac{|a_1|\,\cos\,\beta + |a_2|\,\cos\,\gamma\,q^2}{(1+q)^2}
\end{equation}
where $\cos\,\beta = \hat{L} \cdot \hat{S}_1$ and $\cos\,\gamma = \hat{L} \cdot \hat{S}_2$.

As mentioned before, the default spin parameter of SHARK MBHs is $a_1 = a_2 = 0.67$ and there is no information about the directions. We therefore randomly sample unit vectors for both spins and orbital angular momentum, i.e. $\hat{S}_1$, $\hat{S}_2$, $\hat{L}$ over a surface of a sphere. To further justify this choice we note that, as we consider only systems with low $f_{\rm edd}$, we assumed that the spin alignment would be negligible (discussed further in Sec.\,\ref{subsec:dual}). Finally, we can calculate the intrinsic spin-flip angle as:
\begin{equation}\label{eq:spin_flip}
    \cos\,\theta = J \cdot S_1 
\end{equation}

To assess how much varying spins can change the flip angles, we additionally explore the sample with randomly assigned spin parameters (and either pro- or retrograde orbits) for all MBHs. 

For each projected angle $\theta'$ (i.e. randomly rotated $\theta$), we apply additional detectability criteria, in order to evaluate whether the source can indeed be identified as an XRG in observations. Following \cite{Rottmann02} these are: a) $ 20^{\circ}~<~\theta'~<~160^{\circ}$ and b) $i<40^{\circ}$, as jets with large inclinations $i$ will be strongly affected by relativistic beaming and hence more likely to be missed in surveys.

\subsection{Dual-jetted AGN}\label{subsec:dual}

As can be seen from Fig.\,\ref{fig:qs}, the majority of jets in our initial sample can be found in a pair, i.e. as dual-jetted AGN. 
This is somewhat contradictory to observations (however scarce) and it is likely associated with our initial selection criteria and limitations of our simulations (especially the resolution). This is due to the fact that we select jetted AGN based on conditions related to parameters which are not followed down to close separations, but are written out at the moment of galaxy merger (e.g. mass and accretion rates). This issue could be fixed with modelling the evolution of the above parameters in post-processing, which we leave for future work.

Dual-jetted sources are poorly understood and most of our current knowledge comes from theoretical studies of cosmological populations of dual AGN at several (hundred) kpc separations, which are strongly limited by their resolution or from hydrodynamic simulations of individual binaries (see \citealp{DeRosa19} and references therein). Nevertheless, two of the relatively well constrained requirements for dual activity are large gas reservoirs and mass ratios close to unity \citep{Rosado14, Volonteri15, Volonteri15b, Steinborn16, DeRosa19, Volonteri22, Izquierdo-Villalba23}. The former is rather intuitive as jets are powered by accretion disks. The latter on the other hand is related to the fact that in the case of minor mergers, the secondary MBH can be efficiently stripped of its accretion disk even before formation of the Keplerian binary, which prevents generation of a jet \citep{Lobanov08, Mezcua11, Volonteri16}.  

Additionally, it has been suggested that fast rotating gas-rich galaxy disks tend to align MBH spins during the DF phase if MBHs accrete 1\%-10\% of their mass during that time \citep{Bogdanovic07}. If spins are indeed aligned, it would mean that gas-rich mergers cannot be related to XRGs. However, one of our conditions for efficient jet launching are very small accretion rates with $f_{\rm edd}<0.02$, which translates to mass gain below 1\% for virtually all sources in our sample.

In order to constrain our final dual-jetted AGN sample given the above arguments we assume that they can be produced only in gas-rich mergers with mass ratios $q>0.3$ (SHARK definition of a major merger). 
This makes the number of dual-jetted AGN in our sample drop by approximately 50\%. Additionally, we assume that sources that were cut at this step are still valid as binaries with only one jetted AGN and so we add them to our sample of $M_{1}$ only jets for which we calculate spin-flip angles (we mark them further as $dual^*$). 

\subsection{Radio luminosities}
In the case of simulated XRGs we calculate their radio luminosity $L_{\rm 1.4GHz}$ in the units of W/Hz as \citep{Amarantidis19}:
\begin{equation}
    \nu L_{\rm 1.4GHz} = A\,\left( \frac{M_1}{10^9\,\rm M_{\odot}} \cdot \frac{f_{\rm edd}}{0.01} \right)^{0.42}\,L_{\rm jet}
\end{equation}
where $\nu$ is the radio frequency, $A=1.3\cdot10^{-7}$ is the normalisation factor and $L_{\rm jet}$ is the luminosity of the jet given by:
\begin{equation}
    L_{\rm jet} = 2\cdot10^{45}\,\left( \frac{M_1}{10^9\,\rm M_{\odot}} \cdot \frac{f_{\rm edd}}{0.01} \right)\,a^2\, [{\rm erg\,s}^{-1}]
\end{equation}

Radio luminosity of the observed XRGs is calculated from their measured flux density $S_{\rm 1.4GHz}$ and spectral index $\alpha$ via \citep{Bhukta22}:
\begin{equation}
    L_{\rm 1.4GHz} = 4\pi\,d_{\rm L}^{2}\,S_{\rm 1.4GHz}\,(1+z)^{\alpha -1}
\end{equation}

\subsection{XRG: results and discussion}\label{subsec:xrg_results}

\begin{figure}
	\includegraphics[width=\columnwidth]{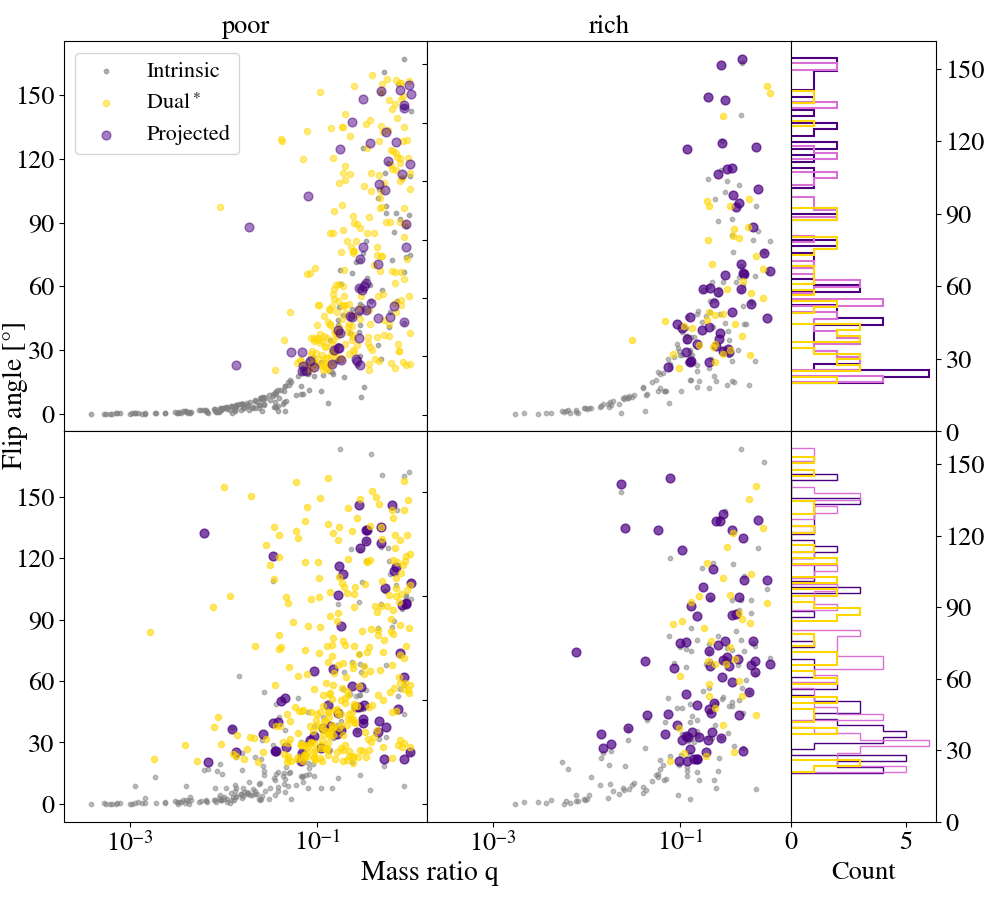}
    \caption{Spin-flip angles as a function of mass ratio. Default spins a=0.67 are shown in the top panel and random spins in the range of 0-1 are shown in the bottom. Grey colour refers to the intrinsic angles, dark purple shows detectable sources given projected angles that fall within our detection criteria and gold corresponds to sources that were initially identified as dual-jetted AGN.}
    \label{fig:angles}
\end{figure}

Fig.\,\ref{fig:angles} shows the intrinsic and projected spin-flip angles as a function of mass ratio calculated only for not-stalled binaries (merger is necessary in this scenario). The top panel corresponds to default spin parameters of $a=0.67$, while the bottom one uses the sample with random spins in the range of 0 to 1. In the case of default spins, only mergers with $q > 0.1$ can produce spin-flips with angles larger than $20^{\circ}$. If we consider random spin parameters, a small fraction of lower mass ratio binaries may produce relatively large flip angles provided that the initial $M_1$ spin is small. Both spin distributions seem to favour rather moderate flip angles with median values of $58^\circ$ and $69^\circ$ for fixed and random spin parameters, respectively. This is in line with observations of XRGs whose angles are found mostly between $50^\circ$ and $90^\circ$. Dual-jetted AGN on the other hand would have a random angle distribution due to our assumptions (both MBH spins and orbital angular momentum are sampled randomly). We note however, that there could be some level of spin alignment during binary migration within the remnant (as noted in Sec.\,\ref{subsec:dual}) as well as due to geometry and dynamics of galaxy mergers. We leave further consideration of this matter to future work.   

\begin{figure}
	\includegraphics[width=\columnwidth]{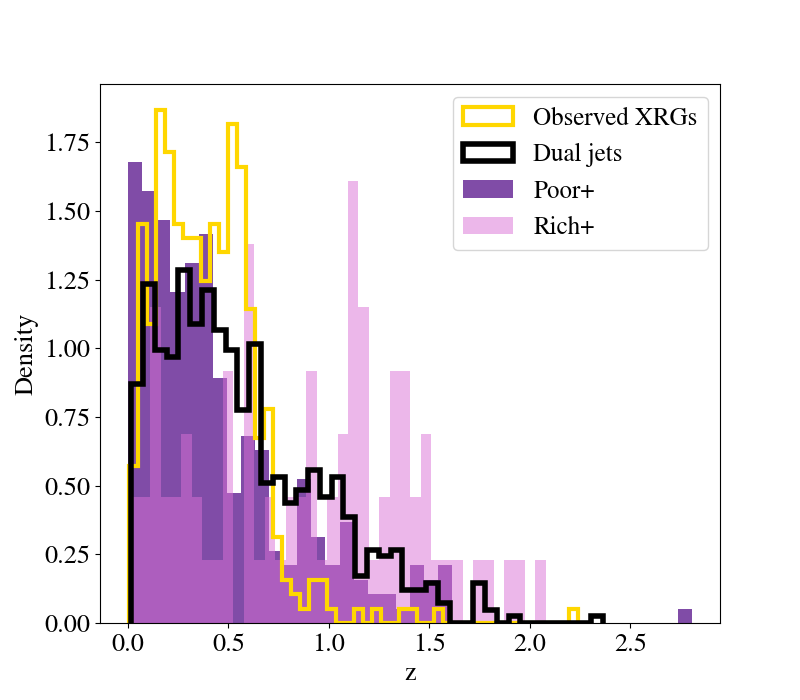}
    \caption{Comparison of observed and calculated XRG redshifts. The shaded pink and purple areas ("Poor+" and "Rich+") correspond to new $M_1$ only jets after including sources initially identified as duals ($dual^*$). Dual-jetted AGN are marked with thick black line and observations with gold.}
    \label{fig:comp_obs}
\end{figure}

In Fig.\,\ref{fig:comp_obs} we show the comparison of observed XRG redshits (for the total number of 432 sources for which the measurement was available) with our estimates. The two distributions shown with shaded pink and purple areas refer to $M_1$ only jets replenished with sources that were initially identified as dual-jetted, but after applying more rigorous criteria were redefined as $M_1$ only jets with an inactive $M_2$. Dual-jetted AGN are shown with a thick black line and the observations are marked with gold. 

The vast majority of observed XRGs is located at small redshifts $z<1$. In the first place, this is a result of severe selection effects and properties of radio emission. Most of the high-z detections correspond to compact quasar images, mainly because a) lobes fade rapidly with distance as the observed frequency is (1+z) times smaller than the emitted one (high frequency emission comes mostly from the jet) and b) high density of cosmic microwave background photons at higher z act as an effective coolant and diminish radio emission \citep{Rottmann02,Diana22, Zhang22}. Furthermore, it is still not clear whether the evolution of RLAGN with jets and non-jetted, radio quiet AGN is a redshift-dependent process and if these two classes of objects follow similar evolutionary paths \citep{Sbarrato21, Sbarrato21b}. In other words, no high-z detections of XRGs can be either associated with observational biases or a lower number of jetted AGN at earlier cosmic times. 

When considering the theoretical population of XRG sources derived in our calculations, we can see that both the twin AGN and gas-poor spin-flip scenarios reproduce the observed redshift distribution well. On the other hand, $M_1$ jets from gas-rich mergers have a relatively flat redshift distribution with a small peak at $z>1$ which is inconsistent with observations. For both scenarios, spin-flip and twin AGN, the maximum redshift of simulated XRGs is at about $z\sim2.5$. These results show that low z of observed XRGs can be associated with intrinsic properties of the sources rather than observational biases if indeed they originate from MBH binaries or their mergers. 

\begin{figure}
	\includegraphics[width=\columnwidth]{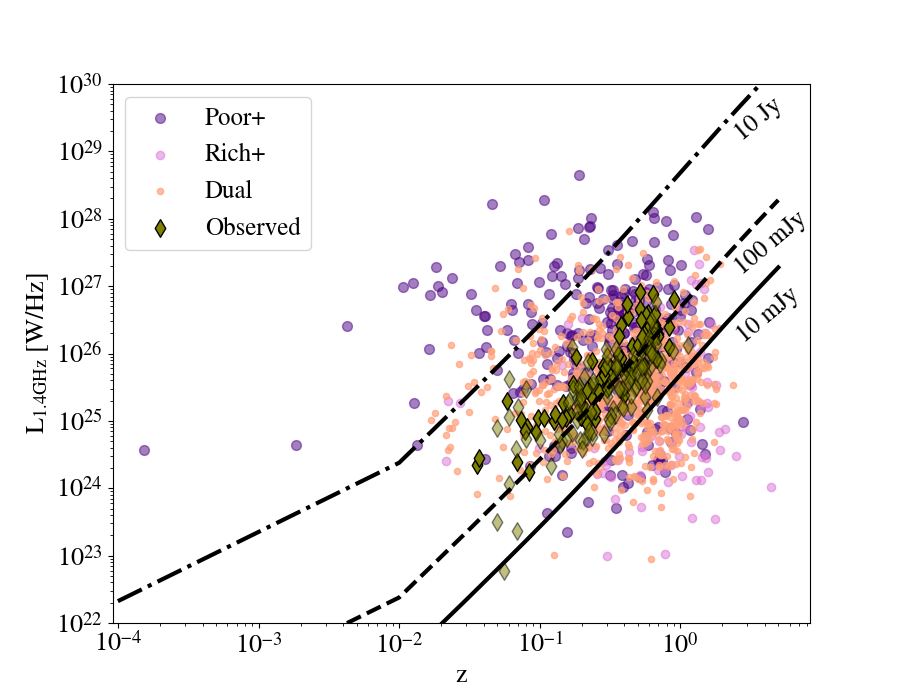}
    \caption{Radio luminosity of the observed (green diamonds) and simulated XRGs. The black lines mark three flux density levels assuming a spectral index of $\alpha\,=\,-0.8$, which is characteristic to synchrotron emission.}
    \label{fig:Lrad}
\end{figure}

To gain further insight, in Fig.\,\ref{fig:Lrad} we show a comparison of radio luminosities $L_{\rm 1.4GHz}$ between observed (olive diamonds) and simulated XRGs (coloured circles) as a function of redshift. The three black lines mark essential flux density levels with the lowermost corresponding to typical XRG detectability threshold. The majority of observed population of XRGs has luminosity in the range of $10^{24}\,-\,10^{27}$\,W\,Hz$^{-1}$ at 1.4\,GHz, which corresponds to flux densities between 10 and a few hundred mJy. It is clear that the majority of simulated XRGs also lie within this range, however there are a few subtleties, which are summarised in Tab.\,\ref{tab:flux}. A substantial fraction of XRGs from gas-poor mergers and duals have flux densities consistent with observed sources. In the case of gas-rich $M_1$ jets, nearly 60\% have flux densities below 10\,mJy which may be attributed to their higher redshifts, smaller MBH masses and hence lower luminosities.  
We also find that 20\% of gas-poor mergers result in overluminous sources which are related to MBHs with $M\sim10^{10}\,\rm M_{\odot}$ at $z\,<\,0.3$. Generation of such overmassive objects in SHARK is an issue likely resulting from a too simplistic treatment of MBH seeding and growth (in fact MBH population is only scaled to $M_{\rm BH}-M_{\rm bulge}$ relation) and should be resolved with more sophisticated models, potentially also including MBH binary evolution. 

\begin{figure}
	\includegraphics[width=\columnwidth]{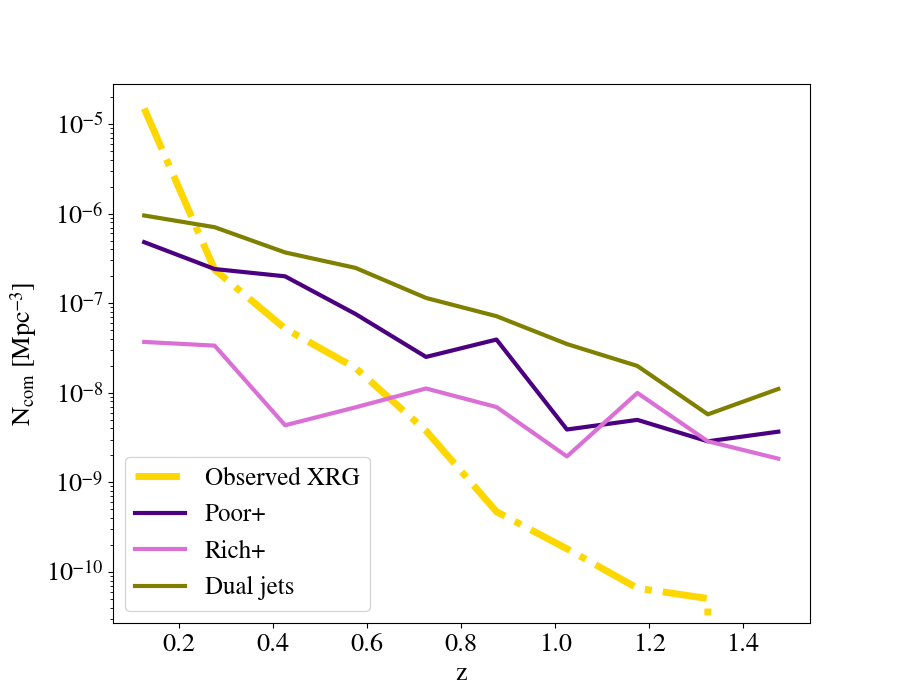}
    \caption{Comoving number density of observed and simulated XRGs as a function of redshift. For the latter, only detectable ($\mathrm{10\,mJy\,<\,S\,<\,10\,Jy}$) sources were plotted.}
    \label{fig:Ncom}
\end{figure}

Lastly, Fig.\,\ref{fig:Ncom} shows the comoving number density $N_{\rm com}(z)$ of observed and simulated XRGs (number of sources $N$ per unit comoving volume per redshift bin) which can be written as:
\begin{equation}
    N_{\rm com}(z) = \frac{N(z)}{V_{\rm c} \Delta z }
\end{equation}
Here, we take into account that radio sources fade away in time and weight comoving densities by probability of observing the source as an XRGs defined as $P = t_{\rm fd}/{\rm d}t$, where ${\rm d}t$ is the width of redshift bin in units of time and $t_{\rm fd}$ is the fading time given by \citep{Rottmann02}:
\begin{equation}
    t_{\rm fd} = 1.56\cdot10^3\,\frac{\sqrt{B}}{B^2+B^2_{\rm IC}}\,\frac{1}{\sqrt{\nu_{\rm b}(1+z)}}\,[{\rm Myr}]
\end{equation}
where $B=2\,\mu$G is the source magnetic field, $B_{\rm IC} = 3.25(1+z)^2$ is the field strength of the inverse-Compton scattering due to cosmic microwave background and $\nu_{\rm b}=1.4$\,GHz is the break frequency. 

Observed XRGs are characterised by a much steeper slope than any kind of simulated ones, which cannot be explained solely by observational limitations as we plotted only detectable sources (with $\mathrm{10\,mJy\,<\,S\,<\,10\,Jy}$). We note however that rarity of these objects together with relatively low sensitivity of large sky surveys may still play a key role, i.e. many XRGs may still remain undetected. A few more factors worth consideration include: a) environmental effects, b) AGN variability and overall uncertainties of jet launch and propagation and c) stronger projection effects. In the case of a), the number of sources we can observe and the slope of Ncom(z) depends on the fading time of the radio emission which has been suggested to decrease with increasing density of the environment due to larger magnetic fields in the clusters of galaxies \citep{Rottmann02}. Further, environmental effects may also affect jet propagation and our rather simple assumptions of jet production could also result in overestimation of the number and strength of radio lobes. The last point c) likely contributes the least to the overall uncertainty, however varying jet efficiencies could result in large disproportion of the two jets enhancing the projection effects on detectability. We are currently unable to specify the true reason of apparent overdensity of simulated XRGs with respect to the observed ones and resolving the issue will require further studies. 

\begin{table}
	\centering
	\caption{Simulated XRGs categorized by flux density levels. Detectable: 10\,mJy\,<\,S\,<\,10\,Jy, underluminous: S\,<\,10\,mJy, overluminous: S\,>\,10\,Jy. As MBH binary spins and orbits, as well as projection angles are sampled randomly, values below correspond to medians calculated over 10 independent runs.}
	\label{tab:flux}
	\begin{tabular}{lc c c} 
		\hline
                       & Poor+ & Rich+  & Dual    \\
		\hline
		Detectable     & 69\%   & 39\%  & 72\%  \\
		Underluminous  & 11\%   & 59\%  & 24\%  \\
		Overluminous   & 20\%   & 1\%   & 3\%   \\
		\hline
	\end{tabular}
\end{table}

\section{Summary and Conclusions}\label{sec:conclusions}
We have presented the study of multi-messenger signatures of MBH binaries and their mergers. The main focus of our work was to: a) estimate the amplitude of the GWB produced by a cosmological population of close binary systems and b) find whether these very same systems can be related to X-shape radio galaxies, i.e. we considered the spin-flip and twin AGN scenarios. We started with modelling MBH binary evolution after the merger of host galaxies in post-processing of the initial populations returned by a semi-analytic model of galaxy formation SHARK. Further, we used the evolved binaries to calculate the GWB and to find and characterise their jet emission. In the case of the latter, we searched for single and double-jetted sources, and compared their properties and occurrence rates with the observed population of XRGs (specifically spin-flip angles for single jetted AGN after their merger, redshift and luminosity distributions and comoving number densities). 

The main conclusions of our work can be summarised as follows:
\begin{itemize}
    \item Accounting for MBH binary evolution within the merger remnant galaxy due to dynamical friction, viscous drag, stellar scattering and emission of gravitational waves results in 40\% of mergers leading to close MBH binary formation out of which 66\% end with MBH coalescence at $z\ge0$ (or 26\% out of all galaxy mergers). 

    \item The estimated GWB amplitude at the reference frequency $f_0=1\,{\rm yr}^{-1}$ is in the range of $A_{\rm yr^{-1}} = 1.20\cdot10^{-15} - 1.46\cdot10^{-15}$ for delayed (accounting for binary evolution timescales) and non-delayed (assuming all galaxy mergers lead to MBH coalescence) mergers, respectively. Small difference between the two amplitudes is a result of optimistic stellar scattering model with a full loss-cone and the fact that only 11\% of all galaxy mergers contribute significantly to the GWB (i.e. have MBHs with masses $M > 10^7$ and redshifts $z<2$). The calculated GWB amplitude is 50\% lower than the signal detected recently by all major PTAs ($A_{\rm PTA} = 2.04\cdot10^{-15}\,-\,2.5\cdot10^{-15}$). 

    \item In the case of the spin-flip scenario, where pre-merger jet changes its direction due to MBH coalescence, we find that the angles between the old and new jets calculated based on our binary sample have median values of $58^\circ$ and $69^\circ$ for MBHs with fixed ($a=0.67$) and random spin parameters, respectively. The result is consistent with the observed population of XRGs whose angles are mostly found in the range of $50^\circ$\,-\,$90^\circ$. This means that the spin-flip scenario provides a physical explanation for the XRG angle distribution. On the other hand, dual-jetted AGN in our model would be characterised by random angle distribution and so they would not reproduce the observed population well. We do however expect some level of spin alignment due to interaction with gas and galaxy merger dynamics, and so we cannot fully disprove this scenario and leave quantifying the effect to future study. 

    \item Distribution of redshifts for our simulated population of XRGs from galaxies that are gas-poor (in spin flip scenario) and gas-rich (in dual-jetted AGN scenario) is consistent with observational data. We estimate that merger/binary induced XRGs could be observable up to redshift $z\sim2.5$ (the farthest known XRG was found at $z~=~2.245$; \citealp{Bera20}) and it largely depends on MBH binary evolutionary timescales. If new, high sensitivity radio observations reveal a substantial population XRGs at larger redshifts it would mean that either binary lifetimes are significantly shorter or at least high-z XRGs are not merger/binary induced. 

    \item Spin-flip induced XRGs in gas-rich galaxies have a rather flat distribution of redshifts with a slight peak at $z>1$ which is inconsistent with observations. We also find that 59\% of them have flux densities below detectability threshold which we set to $S~=~10\,{\rm mJy}$. It might be true that these sources are simply too faint to be currently detected, however their comoving number density is also significantly lower that the observed population and so we find that scenario rather unlikely. 

\end{itemize}

Given the above arguments, we find that spin-flip in gas-poor mergers is a viable candidate for being an XRG progenitor. At the same time, the very same type of sources is also the main contributor to the GWB at the frequency range probed by the PTAs, which possibly opens new avenues for joint multi-messenger studies. In our work, we found that both the GWB and the comoving number density of spin-flip sources at $z<0.2$ are underestimated with respect to the observations and in general, the slope of Ncom(z) for the observed population is much steeper. In the first place, it might be true that not all observed XRGs are genuine sources as most of them are still considered candidates (for a discussion see \citealp{Roberts15}) as well as the signal detected by PTAs will require further evaluation. On the other hand however, we argue that galaxy evolution models are prone to a large number of uncertainties and might need more sophisticated treatment and caution when compared with various observational data which is also often model dependent. Our findings can be therefore verified and explored further in the future by larger parameter space simulations focused on MBH binary evolution, more detailed studies of jet/lobe luminosities and upcoming sensitive radio sky surveys.

\section*{Acknowledgements}
MC is supported by the Polish National Science Center through research grant NR 2021/41/N/ST9/01512.
TB is supported by the International Research Agenda Programme AstroCeNT
(MAB/2018/7) funded by the Foundation for Polish Science (FNP)
from the European Regional Development Fund. TB also acknowledges support from the Polish National Science Center grant Maestro (2018/30/A/ST9/00050).

\section*{Data Availability}
The data directly underlying this article will be shared on reasonable request to the corresponding author. The SHARK model is publicly available and can be retrieved from \url{https://github.com/ICRAR/shark}, while halo and merger trees are available upon request to the SURFs team (surfs@icrar.org). 


\bibliographystyle{mnras}
\bibliography{bibliography}

\bsp
\label{lastpage}
\end{document}